\documentclass{aastex} 
\usepackage{emulateapj5}  

\usepackage{psfig}

\begin{document}

\newcommand{\asec}{\ifmmode ^{\prime\prime}\else$^{\prime\prime}$\fi}
\newbox\grsign \setbox\grsign=\hbox{$>$}
\newdimen\grdimen \grdimen=\ht\grsign
\newbox\laxbox \newbox\gaxbox
\setbox\gaxbox=\hbox{\raise.5ex\hbox{$>$}\llap
     {\lower.5ex\hbox{$\sim$}}}\ht1=\grdimen\dp1=0pt
\setbox\laxbox=\hbox{\raise.5ex\hbox{$<$}\llap
     {\lower.5ex\hbox{$\sim$}}}\ht2=\grdimen\dp2=0pt
\def\gax{$\mathrel{\copy\gaxbox}$}
\def\lax{$\mathrel{\copy\laxbox}$}
\def\bh{black hole}
\def\acn{accretion}
\def\acg{accreting}
\def\ac{accrete}
\def\rd{Di\thinspace Stefano}
\def\gy{galaxy}

\title{A NEW WAY TO DETECT MASSIVE BLACK HOLES IN GALAXIES:
The Stellar Remnants of Tidal Disruption}

\author{R. Di Stefano}
\affil{Harvard-Smithsonian Center for Astrophysics, Cambridge, MA 02138}
\affil{Department of Physics and Astronomy, Tufts University, Medford, MA 02155}
\author{J. Greiner}
\affil{Astrophysical Institute Potsdam, 14482 Potsdam, Germany}
\author{S. Murray, M. Garcia}
\affil{Harvard-Smithsonian Center for Astrophysics, Cambridge, MA 02138}

\begin{abstract} 
We point out that the tidal disruption of a giant may leave a luminous
($10^{35}-10^{39}$ erg/s), hot ($10-100$ eV) stellar core.
The ``supersoft" source (SSS) 
detected by {\it Chandra} at the center of M31
may be 
 such a core;  
whether or not
it is, the observations
have shown that such a core is detectable, even in the center of
a galaxy. We therefore explore the range of expected observational signatures,
and how they may be used to (1) test  
the hypothesis that the M31 source is a remnant of tidal stripping,
and (2) discover evidence of \bh s and disruption events in other
galaxies.   
\end{abstract} 


\section{Tidal Disruption of Giants: The Core Remains}


A star of mass $M_\ast$ and radius $R_\ast$ can be tidally 
disrupted by approaching
within $R_t$ of a black hole (BH). 
\begin{equation}
    R_t=\Bigg( \eta^2 {M_{bh} \over {M_\ast}} \Bigg)^{{1}\over{3}}\, R_\ast,
\end{equation} 
$M_{bh}$ is the BH's mass;
$\eta$ is a parameter of order of unity. 
Let $R_S$ be the BH's Schwarzschild radius.
For $M_{bh}$ \gax\
 $10^8\, M_\odot,$
$R_t < R_{S}$ for most main-sequence stars,
and only tidal disruptions (TDs) of giants lead to 
observable effects.
The rate of such TDs, $\dot N_{G},$  depends on $M_{bh}$, the   
kinematics of stars near the galactic nucleus,
and the mass and age distribution of these stars (see, e.g.,
Magorrian \& Tremaine 1999). Computed
rates are typically in the range $10^{-6}-10^{-4}$ yr$^{-1}$.
A convenient expression is
\begin{equation}
\dot N_{G} \sim 10^{-5} \Big({{L}\over{L_\ast}}\Big)^{1.2} yr^{-1},  
\end{equation}
with $L_\ast =1.8 \times 10^{10} L_\odot;$ 
this was derived by Syer \& Ulmer (1999),
who also provide cautions about the use of
a simple scaling law.
Studies of TD have  
focused on
possible associated accretion events,
lasting for months or decades, with luminosities as
high as $\sim 10^{44}-10^{46}$ erg/s (see, e.g.,
Hills 1975, Lidskii \& Ozernoi 1979,
Gurzadyan \& Ozernoi 1980, Rees 1988,
Loeb \& Ulmer 1997, Ulmer, Paczy\'nski, \& Goodman 1998, Ulmer 1999).
 
\noindent{\bf The Hot Core of a Disrupted Giant:\ } 
The disruption leaves an end-product, 
the
giant's hot dense core,
whose presence, influence,
and observability are the subjects of this paper. 
The core remains hot ($T>10^5$ K) and bright ($L>10^2 L_\odot$) 
for 
$10^3-10^6$ years, thereby providing  
the longest-lasting
signal of a TD.

\noindent{\bf Detectability:\ }
The soft X-ray sensitivities of {\it Einstein}, ROSAT, 
{\it Chandra}, 
and XMM-{\it Newton}
have allowed them to detect supersoft X-ray sources (SSSs) in other galaxies.
SSSs (Greiner et\,al. 1991, Rappaport, \rd\ \& Smith 1994, 
Kahabka \& van den Heuvel 1997)
have luminosities ($10^{35}-10^{38}$ erg/s) and temperatures 
($k\, T \sim 10-100$ eV), comparable to those 
of hot stellar cores.
{\it Chandra's} 0.5--1\asec\ 
angular resolution allows 
individual sources to be resolved, even in the dense central
regions of nearby galaxies. 

\noindent{\bf The Example of M31:\ }
We have studied 8.8 ksec of data collected by Chandra's ACIS-I
detector. To identify SSSs 
we applied a hardness ratio test
to all point sources in a $16'\times  16'$ region centered on
the center of M31, selecting all sources with more than 50\%
of their photons below 0.7 keV. Three sources satisfied this condition.
Remarkably, one of these is coincident with the center of M31.
The luminosity   
is $\sim 10^{37}$ erg/s; the data are consistent with little or
no emission above $1.5$ keV.
The
source appears to be variable; 
data
from earlier missions ({\it Einstein}, ROSAT), e.g.,  
are consistent with the source providing zero flux on one occasion. 
Garcia {\it et al.} (2000) have argued that the density of point X-ray
sources in this field is so low that a chance coincidence
with the center is highly unlikely; this argument is even stronger when
applied to SSSs. 
The observed SSS-like behavior 
is therefore most likely 
related to the environment of the center, presumably to the 
presence of a MBH.
The data do not seem consistent with ADAF disk models
(see, e.g., Garcia {\it et al.} 2000), but
are compatible with the signature of  
a hot stellar core. 
Whether or not the source is a hot core, the
M31 observations establish
that it is possible to detect such an object
in the centers of nearby galaxies.

\section{The Nature of the Cores}

The nature and characteristics of the core depend on the initial
mass, $M_\ast$ of the stripped star, and on its state of evolution at the 
time of disruption. 
 
\subsection{$M_\ast < 8 M_\odot$} 
Tidal stripping brings a premature halt to  
nuclear burning. 
While any residual nuclear burning continues,  
the luminosity remains at the value
it had when the star was a giant. The  radius shrinks, and
the effective temperature increases.
Depending on the core mass, these systems will stay at
$L > 10^2 L_\odot$, and $T>10^5$ K  for
$10^3 - 10^5$ yrs (see, e.g., Bl\"ocker 1995; 
Gorny, Tylenda \& Szczerba 1994).

Nature has provided direct analogs, in the form
of the hot cores at the centers of planetary nebulae.
A notable example is 
1E 0056.8-7154, a planetary nebula system in the Small Magellanic Cloud, with 
$L = 2\,\times\, 10^{37}$ erg/s
and $k\, T = 30-40$ eV
(Wang \& Wu 1992).

The evolution of the hot core is influenced by its mass,  
its state at the time of disruption (e.g., its thermal pulse-cycle
phase), 
 and  the mass of the original star
 (see, e.g., Iben 1995 and references therein). 
 The hot core of a star that 
was disrupted at 
some arbitrary time in its evolution, is descended from
a more massive star than a core of the same mass emerging during the 
final stages of an uninterrupted evolution.
The core of the disrupted star is therefore younger, and is generally
less compact, with  
a significantly higher core temperature; it cools more slowly
than typical post-AGB stars.
(See, e.g., Bl\"ocker 1995;
Gorny, Tylenda, \& Szczerba 1994.)
 Although these results were arrived at
by studying winds, they are expected to be valid for the 
extreme mass-loss scenario associated with disruption.  

The stellar mass function favors stars in this mass range,
and they are therefore expected to be disrupted more frequently than
stars of higher mass. Nevertheless, we go on to consider high-mass stars,
because their disruption may also produce remnants with observable signatures. 

\subsection{$ M_\ast > 8\, M_\odot$}

The stripped core is a helium star (He star).
Two processes (in addition to TD)
may produce He stars.
First,
a very massive star may eject such copious
winds that it effectively ejects its own hydrogen
envelope. Second, a massive star in a binary may
overfill its Roche lobe as it evolves and lose its
hydrogen envelope by either transferring it to
its companion (if the mass ratio is not too
extreme), or losing it during a common envelope episode.
The evolution of isolated He stars and He stars 
in binaries has been well-studied; see, e.g., Maeder (1981),
Delgado \& Thomas (1981), Habets (1986, 1987), 
de Loore \& De Greve (1992), Reese (1993), 
Woosley, Langer, \& Weaver (1995).
The characteristics of each stage of the He star's
evolution depend on many factors, including the mass
of the star when its envelope was lost, and the continuing
rate of mass loss to winds. 
A common feature of these evolutions is, however,
that for a wide range of
system parameters, there is a time interval 
lasting up to several million years, during which the
star is both very luminous ($10^{38}-10^{39}$ ergs/s) and
very hot, with $k\, T$ near $10$ eV. 
Characteristic evolutionary sequences include an epoch of helium
burning, a shorter epoch during which heavier elements are burned,
and, in some cases, a Type Ib or Type Ic,
hydrogen-poor, supernova.

Typical temperatures are lower, and 
typical luminosities higher than those considered in \S 2.1.
Another distinction is that the 
core may not be hottest and brightest 
immediately after the TD. This is because the
further evolution of the He star will typically increase
its luminosity and alter its temperature; in addition,
winds that can block emergent soft X-rays
tend  to decrease with time.

Wolf-Rayet stars may be the closest well-observed analogs of the
He-star remnants of TDs.
They tend to eject such significant winds, however,
that their appearance may be  dominated by the interaction of 
this surrounding matter with radiation from the He star. 
Perhaps because of this, no known He star
has a ``supersoft" signature.
It is possible, however, that some of the known SSSs,
whose fundamental nature is not yet understood, could be He stars.

He-star remnants of 
TDs may be more likely than other He stars to be observed as SSSs.  
This is because the BH continues to act as a sort of 
vacuum cleaner for mass outside the tidal radius 
($\sim R_t \, [M_\ast/M_{bh}]^{{1}\over{3}}$). 
Furthermore, since 
self-generated winds tend to decrease with time
(see, e.g., Woosley, Langer, \& Weaver 1995), there
will typically be less and less material within 
the tidal radius, even
as nuclear evolution continues. 
Thus, He stars near MBHs, especially those that
have undergone TD, may provide the  cleanest signatures 
of the underlying properties of the star, distinct from the
 signatures of its interactions with its environment.

\section{Observational Signatures}

\noindent {\bf Luminosity:} 
For both white dwarfs and He stars,
the luminosity of the stripped core should not exceed
 that of typical giant stars. Luminosities in excess of 
$\sim 10^{39}$ erg/s must be due to some alternative or additional
effect.

\noindent {\bf Spectrum:} 
The maximum temperature expected for hot white dwarfs
(He stars) is $\sim 100$ ($15$) eV. Therefore, 
when $N_H$ is low, the X-ray spectrum is dominated
by soft photons. There can also be 
a modest contribution from photons at higher energies. 
First, the spectrum could differ from a simple thermal shape, e.g.
due to opacity effects, as is well known 
 in the case of white dwarf atmospheres 
(see, e.g., Dupree \& Raymond 1983,
van Teeseling et al. 1994, Hartmann \& Heise 1997);
thus, atmospheric effects must be included when fitting the X-ray
spectrum.
 Second, there could be 
contributions from a corona, an infall disk within the stellar core's
Roche lobe, or shock heating
of
the ambient medium.
Finally, it may not be possible to resolve the hot core and the
MBH, which should be accreting matter at a low level, e.g.,
from stellar winds. 
Note that any radio emission is more likely to be originating from the
region around the MBH than from the hot remnant of TD. 

\noindent It is important to
look for distinctions between the signature of tidally stripped remnants 
and other possible explanations for  SSS signatures.
 One distinction, e.g., between SSSs in binaries and stripped cores,
should be that the distinctive disk signatures of the former 
(Popham \& Di\thinspace Stefano 1996)
should be missing in the latter. 


\noindent {\bf lonization:\ }
Even if the total flux from a stripped core over its lifetime
(which, for the purposes of ionization signatures, may be defined to be the
time interval during which $T>50,000$ K) is smaller
than that from the accretion flare event, the number of high-energy
ionizing photons it emits may be significantly larger. 
The distinctive signature 
is photoionization of high-energy states.


\noindent{\bf Time Variability:\ }
Intrinsic variability may occur at many time scales.
Here we concentrate on variability due to absorption of
soft X-radiation by a varying amount of matter lying
between the source and our detectors.  
Figure 1 demonstrates that the decline in count rate 
with increasing column density can be dramatic.

\smallskip
       \psfig{figure=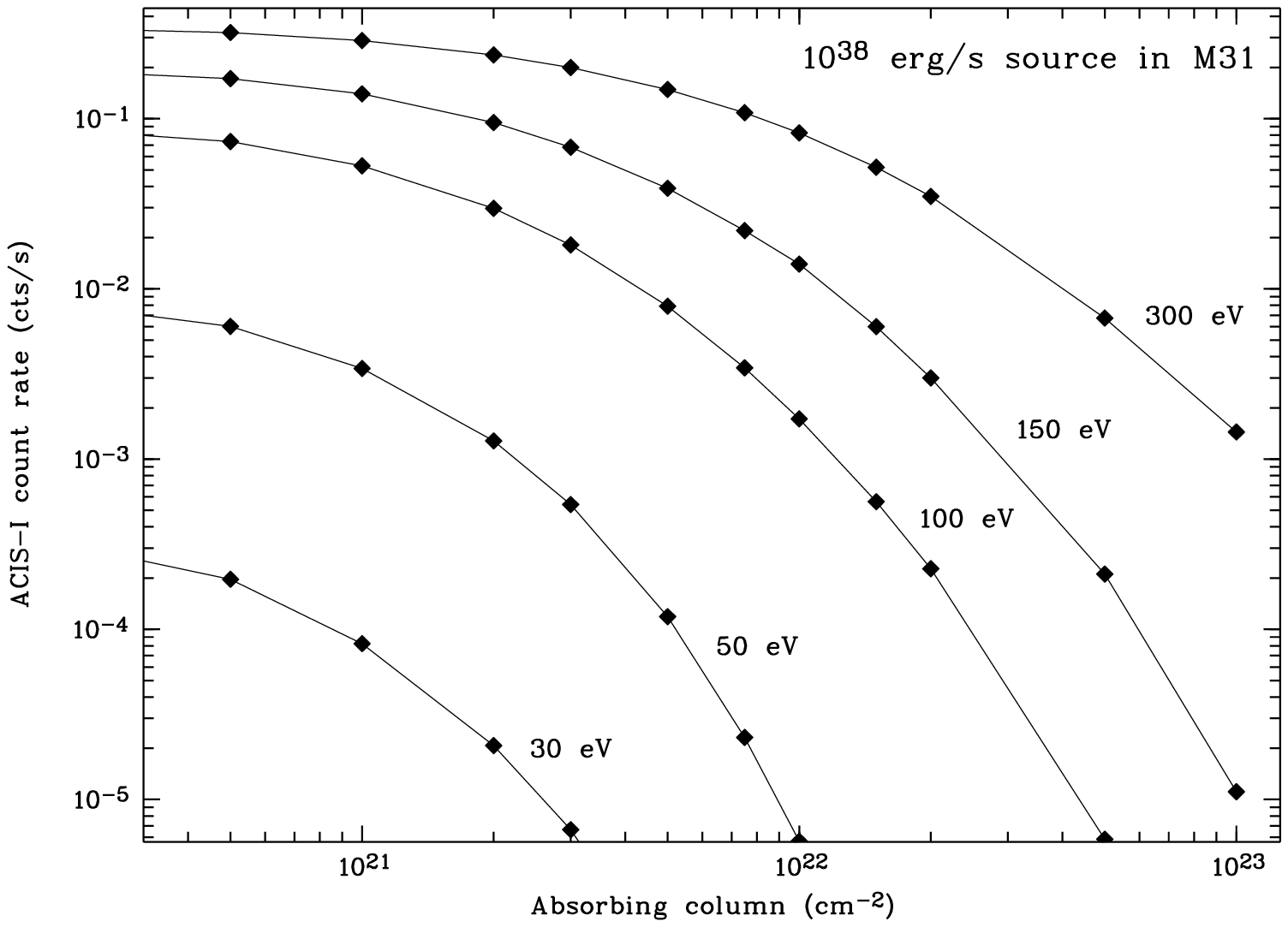,width=8.2cm}
      \noindent{\sl Figure 1: {\it Chandra} ACIS-I detector count rates of a
      black body radiator with
      a bolometric luminosity of 10$^{38}$ erg/s in M\,31 for different 
      temperatures (solid lines) and depending on
      the intervening absorbing column.}
\bigskip\bigskip

If the stripped core orbits the MBH, the flux we receive may
decrease as it travels
behind gas in the disk around the MBH.
If the mass and radius of the star were $M_{\ast}$ and $R_{\ast}$
before the TD, and if the orbital separation
between the core and the MBH is presently equal to the separation
at which the star filled its Roche lobe, then
\begin{equation}
P_{orb} = 0.32\, years\ \eta\,
\Bigg[{{R_{\ast}}\over{100\, R_\odot}}\Bigg]^{{3}\over{2}}
\Bigg[{{M_\odot}\over{M_{\ast}}}\Bigg]^{{1}\over{2}}
\end{equation}

If the core escapes from the vicinity of the MBH, it could
also exhibit variability as it passes behind gas clouds. Making a set
of simple assumptions, we estimate the amount of time, $\tau,$ it takes
for the hot core to pass behind a cloud with column density $N_H,$
located a distance $a$ from the massive \bh .\footnote{
Consider a cloud with mass $m$ located a distance $a$ from the
MBH. If the cloud is corotating, the largest volume it  could
have is roughly equal to the volume of a sphere whose
radius is the effective Roche lobe radius. This allows us to
relate the mass of the cloud to its radius $R$, its distance $a$ from the 
MBH, and to $M_{bh}$. Assume the cloud is composed
of hydrogen and that we can approximate an average value of
 $N_H$ by the number of hydrogen atoms
divided by the square of the cloud radius. This allows us to relate $R$ to $N_H.
$
Finally, we use $\tau = R/v$.}
\begin{equation}
\tau= {{3.5\,  days}\over{v_{150}}}\,  \xi \, 
\Bigg[{{N_H}\over{N_{H,O}}}\Bigg]
 \Bigg[{{a}\over{pc}}\Bigg]^3
\Bigg[{{3\times 10^7\, M_\odot}\over{M_{bh}}}\Bigg], 
\end{equation}
where $\xi$ is of order unity, $v_{150}$ is the transverse speed
in units of $150$ km/s, and $N_{H,0}=1.9\times 10^{22}$.
The actual duration of obscuration may be shorter or longer, depending
on the geometry of the cloud and the track of the hot remnant.

\noindent{\bf
Doppler shift:\ }
The orbital speed of the star in a circular orbit around a
$3 \times 10^7 M_\odot$ \bh ,
with orbital separation, $a,$ such that a $100\, R_\odot$ star
fills it Roche lobe, is $\sim 5\%\, c;$ the speed of the hot remnant
could be greater.
 Thus, the most direct confirmation that an observed SSS is
a tidally stripped remnant, would be a periodic Doppler shift
compatible with an orbit small enough to allow 
TD of the progenitor star.
Significant blue or red shifts
could possibly be inferred from spectral fits.
Even if the remnant does not orbit the MBH,
it might be possible to measure its velocity and
establish the likelihood that it is escaping from the central
gravitational potential.

\noindent{\bf  Spatial Distribution:\ } If the hot stellar core escapes from
the region around the MBH , traveling along a direction perpendicular
to our line of sight, it could traverse a considerable distance before
it dims and cools. If the average speed is $\sim 10^8$ cm/s, the core
could travel $\sim 1$ ($0.1$) kpc in $10^6$ ($10^5$) years.
If the product of the TD rate and
average hot-core lifetimes is larger than unity, we might expect to see
several hot cores ringing the MBH, with the oldest cores further
away. Any cores located $\sim 1$ kpc from the galaxy center 
would likely be He stars.
    It is interesting to note that the other
    two SSS candidates near the center of M31 are each
    several arcminutes from the galaxy's center, produce count
    rates similar to that of the central SSS,
    and are each considerably softer than the central source.

\noindent{\bf Population Signatures: }
(1) {\it Consistency with the model:\ }
If $N$ M31's SSSs are remnants of TD events, each
represents a much larger population of
giants in the galaxy's central region at the time of the TD.
We must therefore test whether the present-day population
could have evolved from one with the requisite number of giants some
$10^3-10^6$ years ago.
If we determine that a subset of the TD remnants
are He stars, it is likely that there were
a large enough number of centrally-located high-mass stars active in
the recent past to lead to some supernovae.
We can ask if the characteristics of the central region (up to
$\sim 1$ kpc from the center) are consistent with the required number
of Type II or Type Ib and Ic supernovae.
The degree and characteristics of the ionization of the central region
must also be consistent with the effects expected from the
 $N$ TD remnants and
the requisite total population of high-mass stars.
It may be instructive to consider the population of
He stars near the center of the Milky Way, which is much better studied.
This population 
is large and includes members with apparently unusual
    properties.  (See, e.g., Tamblyn {\it et al.} 1996, Najarro
    {\it et al.} 1994 and references therein.)
    Unfortunately the large column density
    toward the center of the Milky Way makes it impossible to study
    any soft X-ray emission from these stars.

(2) {\it Eliminating other possibilities:\ }
Population studies of the region around the galactic center
can lead to estimates of the numbers of post-AGB stars and
also of X-ray binaries of various types. The physical processes that
give rise to each of these objects also may give rise to
SSSs that have nothing to do with TD.
We therefore need:  (1) observations of the central population,
(2) theoretical estimates of the number of SSSs not due to TD 
expected, and (3)
comparisons with studies of populations located away from
the central region.
These should allow us to estimate the probability that soft sources observed
near the center have their genesis through processes that are
ordinary and expected parts of stellar and binary evolution.

\section{Other Galaxies} 


Radiation from a hot stripped core can 
be detected in other galaxies if 
(1) the  count rate is high enough, and (2) the source can
be resolved.

\noindent {\bf Count Rate:\ }
The count 
rate from the soft X-ray source in the center of
M31 (Garcia {\it et al.} 2000) was $\sim 10$ counts/ksec. 
Based on spectral fits to nearby (brighter) sources,
the absorption was estimated to be $2.8 \pm 1.0 \times 10^{21}$ cm$^{-2}$.
The number of counts expected from an $E$ ksec exposure of
 a source with  similar spectrum,
located a distance $D$ from us,
is approximately
\begin{equation}  
N_{\gamma} = 10\, {counts} \ \times E \, \times \Bigg({{Mpc}\over{D}}\Bigg)^2  
\times f\Bigg({{N_H}\over{2.8 \times 10^{21}}}\Bigg).   
\end{equation}  
$N_H$ is the column density between us and the new source. 
For face-on or early-type
 galaxies $N_H$ could be smaller than $2.8 \times 10^{21}$,
increasing the count rate in a way that depends on the spectrum, but
which is likely to be described by a non-linear monotonically decreasing
function, $f({{N_H}\over{2.8 \times 10^{21}}}).$ (See Figure 1.)

A $50$ ksec exposure could therefore yield 
20 counts from a hot stellar core an in M31-like galaxy as far away as 5 Mpc; 
for galaxies with less internal absorption, 
the count rate would be higher.
Should the photons' energies be clustered  
below 1 keV, we would have convincing
evidence for a centrally located soft X-ray source.
XMM-{\it Newton}, with its higher
effective area, could achieve similar numbers of counts 
for galaxies up to 10 Mpc away. 
There are about 200 (600) galaxies within 5 (10) Mpc listed in Simbad and
the Principal Galaxies Catalog (PGC).
The galaxy type or orientation could allow
the detection of  
soft central X-ray sources in roughly $1/3$ of these galaxies.  
 
\noindent{\bf Spatial Resolution} In M31 there are $4$ other X-ray sources
 within
$\sim 10''$ of the nucleus. 
If the central density  
of X-ray sources in other galaxies is similar,
 we might not be able to resolve the central source in
galaxies farther from us than $2-3$ Mpc.
Note, however, that, if the soft central source is variable, as the source in
M31 appears to be, it may be possible to successfully apply to
X-ray searches for central sources, the same difference imaging techniques
that have been applied to optical searches for supernovae  and microlensing.
If the hard X-radiation is steady, or varies in a way well-known from
studying other X-ray stars, then we may be able to infer the
existence of the soft central source. XMM-Newton, which has $4-5$ times 
{\it Chandra's} effective area at low energies, will make such differencing
studies possible for M31-type galaxies at least as far away as $5$ Mpc.
This is particularly so if simultaneous
 observations with {\it Chandra} can resolve some sources and
help to  
determine their spectral characteristics.

\noindent{\bf Statistical Tests: \ }
Statistical questions can be posed when we study a population of galaxies.
For example, are galaxies for which there is good evidence (perhaps
from dynamical studies) for a MBH more likely to house a 
centrally located population
of soft X-ray sources, than galaxies for which the dynamical studies
do {\it not} indicate the presence of a MBH?

\section{Conclusion}

When a giant is tidally disrupted, 
 the stripped core may provide
a hot and bright signature for $10^3-10^6$ 
years.  
Detecting such cores  and relating them to TD 
events is challenging. 
Fortunately, the present generation of instruments, including
{\it Chandra,} XMM-{\it Newton,}  and  HST, 
promise opportunities for progress.

This new method of studying MBHs
complements
searches for flare events that may be due to TDs.  
The latter must of course be used for distant galaxies,
and there is a growing body of evidence for luminous UV and
X-ray events lasting for months to years, which may be
consistent with the accretion of the envelopes of tidally-disrupted 
stars.~\footnote{Examples: the
UV flare in NGC 4552 (Renzini {\it et al.} 1995)
the $\sim 5\times 10^{43}$ erg/s outburst in IC 3599 (Brandt et al. 1995,
Grupe et al. 1995),
the $\sim 2\times 10^{43}$
erg/s outburst in  NGC 5905 (Bade et al. 1996, Komossa \& Bade 1999),
the $>\, 9\times 10^{43}$ erg/s flare
in RXJ\,1242.6-1119 (Komossa \& Greiner 1999), the
$\sim 10^{44}$ erg/s flare in RXJ\,1624.9+7554 (Grupe et al. 1999), the
$\sim 10^{44}$ erg/s flare in RXJ\,1331.9-3243 (Reiprich \& Greiner 2000), or
the $\sim 2\times 10^{44}$ erg/s
flare in RX J1420.4+5334 (Greiner et al. 2000).}
Such events are so rare, however, that we would
be lucky to detect one in the near future  among the small number
of galaxies within  $5$ or $10$ Mpc. In nearby galaxies, however,
the long-lived signatures of the stripped cores may  allow us
to infer that TDs have occurred. The two complementary
modes of study should help us to better understand the 
frequency and consequences of tidal disruption events
and to use them as probes of massive black holes
in the center of galaxies.

\end{document}